\documentclass[aps,prl,twocolumn,showpacs]{revtex4}

\usepackage{graphicx}
\usepackage{latexsym} 
\usepackage{amsmath,amscd}
\usepackage{times}

\begin{document}

\title{\bf 
Coil-Globule transition of a single short polymer chain - an exact enumeration study}
\author{M. Ponmurugan$^1$, S. L. Narasimhan$^{2,*}$, P. S. R. Krishna$^2$ and K. P. N. Murthy$^{1,3}$}
\affiliation{
$^1$ Materials Science Division, Indira Gandhi Center of Atomic Reasearch, Kalpakkam - 603102, Tamil Nadu, India.\\
$^2$ Solid State Physics Division, Bhabha Atomic Research Centre, Mumbai - 400085, India.\\
$^3$ School of Physics, University of Hyderabad, Central University P.O, Gachibowli, Hyderabad - 500046, Andhra Pradesh, India.}

%\date{} 

\begin{abstract}

We present an exact enumeration study of short SAWs in two as well as three dimensions that addresses the question, `what is the shortest walk for which the existence of all the three phases - coil, globule and the {\it theta} - could be demonstrated'. Even though we could easily demonstrate the coil and the globule phases from Free Energy considerations, we could demonstrate the existence of a {\it theta} phase only by using a scaling form for the distribution of gyration radius. That even such short walks have a scaling behavior is an unexpected result of this work.

\end{abstract}
\maketitle

Proteins are biologically active, long chain heteropolymers that exist either in the collapsed (native folded or molten globule) phase or in the extended coil phase, depending on their environmental conditions ~\cite{KS}. Simple lattice models ~\cite{CD} have been devised for understanding the role of energy-entropy balance in realizing these phases. In fact, even short lattice chains consisting of only sixteen monomers or less have been shown ~\cite{OC} to be long enough to throw light on the various phases of proteins. These chains are basically Self Avoiding Walks (SAW) on a regular lattice, modeling homopolypeptides with appropriately parametrized intrachain interactions.   

Long chain molecules, modeled as SAWs, are known to exist in one of the three conformational phases ~\cite{flory} - namely, extended coil phase, {\it theta} phase and the collapsed phase - characterized by distinct values of the universal size exponent ~\cite{deGennes}. For example, in two dimensions, the size exponent $\nu = 3/4$ (extended phase), $4/7$ ({\it theta} phase) and $1/2$ (collapsed phase) respectively. It is again the change in the energy-entropy balance that brings about transition from one phase to the other. 

In analogy to the lattice protein models, we may ask what is the shortest SAW for which the existence of all these phases can be demonstrated, eventhough a clear-cut distinction between these phases is ascribed  only to asymptotically long ($N\to \infty$) walks. Moreover, the possibility of experimentally observing conformations of a single chain, which is of finite length, and locating the coil-globule transition ~\cite{UY} begs the question whether the {\it theta} phase of a very short chain can be identified. 

In this brief report, we first demonstrate coil-globule transition for a short SAW consisting of twenty steps or less by exact enumeration of its conformations on two dimensional as well as three dimensional lattices. Then, using the scaling form for the distribution of gyration radius ~\cite{victor}, we show that {\it theta} phase can be clearly identified only for walks having a minimum of sixteen steps. The applicability of scaling ideas to such short walks is indeed surprising. 
 
In order to study the thermodynamic properties of SAW, we need to assign {\it energy} to any given conformation. A standard way of doing this for a lattice SAW is to assign a quantum of energy, $\epsilon$, to each pair of non-bonded nearest neighbor sites that belong to the walk ~\cite{deGennes}. So, a conformation with $n$ such pairs of sites, also called {\it contacts}, has energy $E = n\epsilon$. At any given inverse temperature, $\beta$, the canonical {\it Partition Function} for $N$-step SAWs can be written as
\begin{equation}
C_N(\beta) = \sum _{n=0}^{n=n_X(N)}e^{-n\epsilon \beta}c_N(n)
\label{PF1}
\end{equation}
where $n_X(N)$ is the maximum number of contacts an $N$-step walk can have, and $c_N(n)$ is the number of $N$-step SAW conformations with $n$ contacts. It is clear that conformations with a given number of contacts will acquire more (less) weights if $\epsilon$ is negative (positive). Since we are interested in coil-globule transition, we can set $\epsilon = -1$ without loss of generality.

We may rewrite the above {\it Partition Function} in the form
\begin{eqnarray}
C_N(\beta) & = & \sum _{n=0}^{n=n_X(N)}e^{-\beta F(n,\beta)}\\
-\beta F(n,\beta) & \equiv & \beta n + ln[c_N(n)]
\label{PF2}
\end{eqnarray}
where $F(n,\beta)$ is the {\it Free Energy} associated with walks having $n$ contacts. Since $c_N(n)$ has a bell-shaped form, for a given value of $\beta$, the Free Energy $F(n,\beta)$ is a minimum at a certain value of $n$, say $n_F(\beta)$, which could in general be a non-integer. 
%In order to locate $n_F(\beta)$ accurately, we may compute $F(n,\beta)$ also for non-integer values of 
%$n$ in the range $[0,n_X(N)]$ by obtaining interpolated values of $c_N(n)$.

Another way of estimating $n_F(\beta)$ is to write the Free Energy as an implicit function of the squared end-to-end distance, $r^2$:
\begin{equation}
-\beta F({\bar n}_{\beta}(r^2),\beta) \equiv \beta {\bar n}_{\beta}(r^2) + ln[c_N({\bar n}_{\beta}(r^2))]
\label{FE2}
\end{equation}
where ${\bar n}_{\beta}(r^2)$ is the canonical average of the number of contacts made by all the conformations whose end-to-end distance squared is $r^2$, and $c_N({\bar n}_{\beta}(r^2))$ is the corresponding interpolated value. The estimate of ${\bar n}_{\beta ,F}$ at which the Free Energy is minimum agrees very well with that of $n_F(\beta)$ defined earlier. 

\begin{figure}
\includegraphics[width=3.75in,height=2.75in]{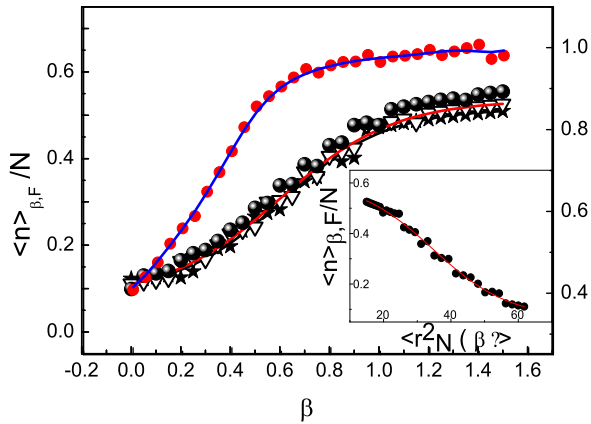}
\caption{Exact enumeration data ${\bar n}_{\beta ,F}/N$ for two dimensional SAWs as a function of $\beta$. (i) Lower one: Square lattice walks of length $N = 16$(filled star), $18$(inverted open triangle) and $22$(shaded filled circle). (ii) Upper one: Triangular lattice walks of length $N = 14$, whose ordinates are marked on the right. INSET: ${\bar n}_{\beta ,F}/N$ as a function of $<r^2>_N(\beta)$ for square lattice walks of length $N = 18$.}
\label{EB2D}
\end{figure}

The temperature dependence of ${\bar n}_{\beta ,F}$ per step, presented in Fig.{\ref{EB2D}} for SAWs on Square as well as Triangular lattices, clearly suggests a transition from a low temperature phase to a high temperature phase. In fact, the specific heat associated with this transition ($C_{sp.ht} = -\beta ^2 \partial {\bar n}_{\beta ,F}/\partial \beta$) is seen to peak at $\beta _{CG} \sim 0.85, 0.52, 1.85$ and $1.35$ on square, triangular, diamond and cubic lattices respectively. What these phases are becomes clear from the inset of this figure which shows a monotonic decrease of ${\bar n}_{\beta ,F}$ per step as a function of the canonically averaged end-to-end distance squared. In other words, the low temperature phase corresponds to compact walks with large number of contacts whereas the high temperature phase corresponds to extended walks with small number of contacts. We observe a similar transition from low temperature compact phase to high temperature extended phase for three dimensional (Cubic and Diamond lattice) SAWs also (see, Fig.{\ref{EB3D}). 

\begin{figure}
\includegraphics[width=3.75in,height=2.75in]{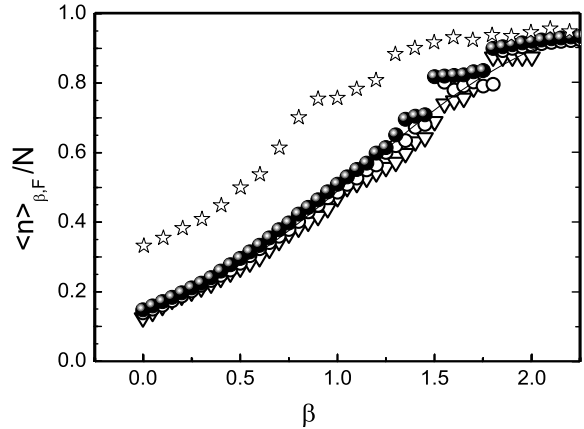}
\caption{Exact enumeration data ${\bar n}_{\beta ,F}/N$ for three dimensional SAWs as a function of $\beta$. Open stars correspond to walks of length $N = 14$ on a Cubic lattice (Data shifted up by $0.25$). The rest of the data correspond to Diamond lattice walks of length $N = 16$(open triangle), $18$(open circle) and $20$(shaded filled circle) respectively.}
\label{EB3D}
\end{figure}

Thus, ${\bar n}_{\beta ,F}$ per step , or equivalently the canonically averaged end-to-end distance squared, seems to be a physical quantity that is responsive to the coil-globule transition. The latter has the advantage of being proportional to the measurable average radius of gyration. But, they provide no evidence for the existence of {\it theta} phase. This leads to the question of how the geometric parameters characterizing a SAW - namely, number of steps in the walk ($N$), number of contacts made ($n$) and the radius of gyration or the end-to-end distance($r_g$ or $r$)- can be put together so as to check whether {\it theta} phase for very short walks is identifiable.

Lhuillier ~\cite{lhuillier} first proposed a scaling form for the distribution of gyration radius and showed that it is related to the Flory free energy for SAW. In fact, radius of gyration is a better geometrical descriptor of a SAW than its end-to-end distance. For example, in the globule or coil regime, smaller (larger) radius of gyration implies larger (smaller) number of contacts; a similar correlation between end-to-end distance and the number of contacts is not so obvious because conformations with different end-to-end distances could all have the same radius of gyration. In the stretched regime, however, radius of gyration and end-to-end distance provide equivalent description because of very few contacts.

Subsequent studies ~\cite{victor} along these lines suggest that the most relevant {\it order-parameter}, $t$, for studying the various phases of SAW is a certain power of the monomer density, namely, $t \equiv \rho ^{1/(\nu d - 1)}$ where $\rho = N/r_g^d$ and $d$ is the dimensionality of space. 
%For example, the $\beta$-dependence of $<t>$ for Square lattice walks of length $N = 20$ ($d =2;\nu =3/$) %is compared, in Fig.{\ref{RsDn2D}, with that of the mean squared end-to-end distance. 
The distribution of $t$ for an $N$-step SAW has a scaling form ~\cite{victor} given by
\begin{equation}
P_N(t) \sim t^c \mbox{exp}(- [a_1t + a_2t^2 + a_3t^{-q}]);\quad q = \frac{\nu - 1/d}{1-\nu}
\label{Pt}
\end{equation}
where $c$ is a universal exponent, related to the susceptibility exponent $\gamma$, having values $-37/32$ and $\sim -1.13$ in two and three dimensions respectively. The exponent $q = 1$ and $\sim 0.617$ in two and three dimensions respectively. The parameters, $a_1, a_2$ and $a_3$, are $\beta$-dependent and, in particular, the value of $\beta$ at which the parameter $a_1$ becomes zero corresponds to the inverse $\theta$ temperature, say $\beta _{\theta}$.

\begin{figure}
\includegraphics[width=.9\hsize]{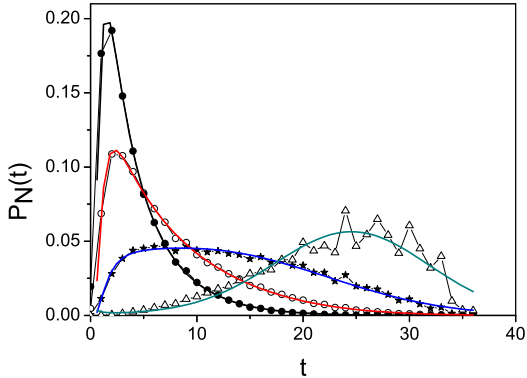}
\caption{Distributions of {\it order parameter} $t = (N/r_g^2)^2$ for Square lattice walks of length $N = 23$ corresponding to $\beta = 0.0, 0.4, 0.8$ and $1.5$ (respectively from left to right). Solid lines are the scaling form Eq. (\ref{Pt}) fit to the the data.}
\label{OPdist}
\end{figure}

\begin{figure}
\includegraphics[width=.9\hsize]{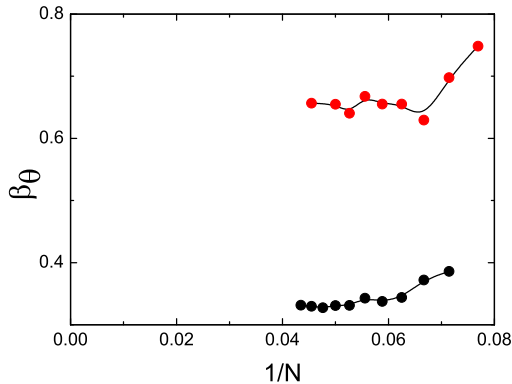}
\caption{$\beta _{\theta}$ values at which the parameter $a_1 = 0$ (Eq.(\ref{Pt})) for various values of $N$ in the range $14$ to $23$ for Square lattice (lower figure), and in the range $13$ to $22$ for Diamond lattice (upper figure).}
\label{betatheta}
\end{figure}

We have shown in Fig.{\ref{OPdist} the distributions, $P_N(t)$, corresponding to inverse temperatures $\beta = 0.0, 0.4, 0.8$ and $1.5$ respectively, obtained from exact enumeration data for $23$-step walks on a Square lattice. By fitting the above scaling form, Eq.(\ref{Pt}), to the histogram data obtained for various values of $\beta$, we find that the parameter $a_1 = 0$ for $\beta = \beta _{\theta} \sim 0.33$. Repeating this procedure, we obtained $\beta _{\theta}$ values for $N = 14$ to $23$ and have presented them in Fig.{\ref{betatheta}; they stabilize around an average value $\sim 0.32$ for $N\geq 16$. Similarly, the $\beta _{\theta}$ values obtained for walks of length in the range $13$ to $22$ on a Diamond lattice (the upper curve in Fig.{\ref{betatheta}) are also seen to stabilize around an average value $\sim 0.65$ for $N\geq 15$. This implies that a {\it theta} phase could be identified for even for such short SAWs.
 
It may be noted that the fit (solid lines in Fig.{\ref{OPdist}) is not as good in the large $t$ or `globule' regime as it is in the small $t$ or `coil' regime. We found this to be the case for Diamond lattice walks also. The reason could be that the walk is too short for the scaling form given by Eq.(\ref{Pt}) to hold good in the entire range. Moreover, the average values of $\beta _{\theta}$, obtainable from the data presented in Fig.{\ref{betatheta}, are different from the corresponding values obtained for asymptotically long walks ($N\to \infty$). Nevertheless, they provide clear evidence for the existence of a {\it theta} phase where the parameter $a_1$ vanishes.

In summary, using exact enumeration data, we have demonstrated the existence of {\it theta} phase besides the usual coil and globule phases even for short SAWs ($N\leq 23$) in two as well as three dimensions. This is analogous to realizing the various phases of proteins by exactly enumerating short lattice (HP) chains with appropriately parametrized interactions. Even though the coil and the globule phases could easily be demonstrated from Free Energy considerations, the existence of a {\it theta} phase could be demonstrated only by using a scaling form for the distribution of the gyration radius. That scaling ideas work for such short walks is a surprising result.

%\begin{figure}
%\includegraphics[width=.9\hsize]{rsq_dens_beta.eps}
%%\includegraphics[width=3.5in,height=2.75in]{rsq_dens_beta.eps}
%\caption{Mean squared end-to-end distance, as well as the average {\it order parameter} $<t> \equiv (N/   %<r_g^2>)^2$, as a function of $\beta$ for Square lattice walks of length $N = 20$. Ordinates for $<t>$ are %marked on the right.}
%\label{RsDn2D}
%\end{figure}

One of the authors (M.P) acknowledges Grant from Council of Scientific and Industrial Research, India: CSIR no.9/532(19)/2003-EMR-I.

$^*$Author for correspondence (slnoo@magnum.barc.gov.in)

\end{document}